\documentclass[12pt]{article}
\usepackage{epsfig}
\usepackage{axodraw}
 
\newcommand{\mrm}[1]{\mathrm{#1}}
\newcommand{\as}{\alpha_{\mrm{s}}}
\newcommand{\qbar}{\mrm{\overline{q}}}

\newcommand{\gtrsim}{\raisebox{-0.8mm}%
{\hspace{1mm}$\stackrel{>}{\sim}$\hspace{1mm}}}

\newlength{\abstwidth}
\newlength{\captivewidth}
\newcommand{\captive}[1]{\rule{5mm}{0mm}%
\begin{minipage}{\captivewidth}%
\caption[small]{#1}\end{minipage}}
 
\begin{document}
 
\sloppy
 
\pagestyle{empty}
 
\begin{flushright}
LU TP 97--12 \\
May 1997
\end{flushright}
 
\vspace{\fill}
 
\begin{center}
{\LARGE\bf Creating a hybrid of matrix elements and parton showers}\\[3mm]
{\Large Johan Andr\'{e}
}\\ 
{\it Department of Theoretical Physics,}\\[1mm]
{\it Lund University, Lund, Sweden}
\end{center}
 
\vspace{\fill}
\begin{center}
{\bf Abstract}\\[2ex]
\begin{minipage}{\abstwidth}
The second-order QCD matrix elements give a very good agreement 
with experimental data on the angular distributions of the four-jet
events in $\mrm{e}^{+}\mrm{e}^{-}$ collisions 
at the $\mrm{Z^{0}}$ resonance energy.
Unfortunately the description of the sub-jet structure is quite poor. 
The alternative approach, parton showers, gives a good description of the 
sub-jet structure but is worse than matrix elements when it comes to the 
angular distributions. Here is presented a hybrid 
between the matrix elements and the parton showers that is intended to
combine the best of the two approaches.
\end{minipage}
\end{center}

\vspace{\fill}
 
\clearpage
\pagestyle{plain}
\setcounter{page}{1}
%
%
\section{Introduction}
%
%
According to the viewpoint of modern particle physics all matter in the 
universe is composed of the particles listed in the table below.
\begin{equation}
\begin{array}{lllll}
&\underline{\mrm{Quarks}} & & \underline{\mrm{Leptons}} & \\
\mrm{Charge} & \mrm{2/3} & \mrm{ -1/3}  & \mrm{-1} & \mrm{0} \\
\underline{\mrm{Generation}}&&&&\\
\mrm{first}
& \mrm{u~(5\cdot 10^{-3})} & \mrm{d~(7\cdot 10^{-3})} & 
\mrm{e~(0.5\cdot 10^{-3})} & \mrm{\nu_{e}~(0)}  \\
\mrm{second}
& \mrm{c~(1.5)} & \mrm{s~(0.2)} & \mrm{\mu~(0.106)} & \mrm{\nu_{\mu}~(0)}  \\
\mrm{third}
& \mrm{t~(170)} & \mrm{b~(5)}~~~~ & \mrm{\tau~(1.784)} 
& \mrm{\nu_{\tau}~(0)}
\end{array}
\label{fermions}
\end{equation}
The approximate particle masses in the parentheses are given in GeV.
All these particles have spin 1/2 and are therefore fermions.
This means that the total wave-function $\Psi(q_{1},\ldots,q_{N})$ of a 
$N$ particle system is antisymmetric under the permutation of two 
identical particles
i.e. $\Psi(\ldots,q_{i}\ldots,q_{j},\ldots) = 
-\Psi(\ldots,q_{j}\ldots,q_{i},\ldots)$~. Thus two identical particles 
$i$ and $j$ can't
have identical quantum numbers $q_{i}=q_{j}$ because this would imply that 
$\Psi=0$. This is the so-called Pauli principle that determines the structure
of the periodic system.
The particles in the second and third generations have the same properties 
as the corresponding particles in the first generation except for the 
masses that increases with the generation number.   
It is only the particles in the first generation that is needed to build the 
matter on earth. 

There are four known fundamental interactions between the matter particles. 
These are the electromagnetic, the weak, the
strong and the gravitational interactions.  For the first three interactions
we have quantum mechanical theories possessing local gauge symmetries, 
in which the forces are mediated
by exchange of virtual particles. The fourth interaction, gravitation, is
currently 
described by Einstein's general theory of relativity, which is a classical
theory. One of the main objectives of current theoretical research is to find
the correct quantum mechanical description of the gravitation.
Fortunately the gravitational interactions between the particles
studied in particle physics experiments are so weak that they can
be leaved out of account. The interactions, their mediating particles
and the matter they affect are listed in the table below. 
\begin{equation}
\begin{array}{lll}
\underline{\mrm{Interaction}} & \underline{\mrm{Particle}} &
\underline{\mrm{Affects}} \\
\mrm{electromagnetic} & \mrm{\gamma} &  
\mrm{all~except~\nu_{e}~,\nu_{\mu}~,\nu_{\tau}} \\
\mrm{weak} & \mrm{W^{+}~~W^{-}~~Z}^{0} & 
\mrm{all~particles} \\
\mrm{strong} & \mrm{g}_{i}~~i=1 \ldots 8 &   
\mrm{all~quarks} \\
\mrm{gravitation} & \mrm{graviton} & \mrm{all~particles}
\end{array}
\label{bosons}
\end{equation}
All these particles have integer spin and are therefore bosons.
The $\gamma$ particle which mediates the electro-magnetic interaction
is a spin 1 particle. The virtual $\mrm{\gamma}$ particle is massive and can
therefore exist in the tree different spin-states $-1$, $0$ and $+1$,
this corresponds to the fact that the classical 
$\mrm{\vec{E}}$ and $\mrm{\vec{B}}$ fields can be
described by three independent parameters. Similar arguments lead to the 
conclusion that the graviton, if it exists, should be a spin 2 particle.
The electromagnetic interaction is described by QED (Quantum Electro Dynamics)
a theory whose predictions agrees extremely well with experimental 
results. This is in part due to the smallness of 
$\alpha=e^{2}/4\pi\varepsilon_{0}\hbar c $
in which the perturbative solutions are expanded.
The weak interaction is described by a theory that combines the electromagnetic
and the weak interaction into a theory of the so called electro-weak 
interaction. In the electro-weak theory the fundamental particles that mediate
the interactions are the B, $\mrm{W^{0}}$, $\mrm{W^{+}}$ and $\mrm{W^{-}}$ 
particles.
They are all massless, a condition that is necessary for a consistent theory. 
The $\mrm{W^{+}}$ and $\mrm{W^{-}}$ particles acquire their masses through 
interaction with the Higgs field that fills all of space. This is analogous
to the repeated absorption and emission of photons e.g. in glass, which
slows down the overall velocity of the photons and thus gives them an
effective mass that is larger than zero. The same Higgs field polarizes the
superposition of B and $\mrm{W^{0}}$ into the physical states 
$\mrm{\gamma}$ and $\mrm{Z^{0}}$, where the $\mrm{\gamma}$ remains massless
and the $\mrm{Z^{0}}$ acquires a mass.

The theory of the strong interaction is described by QCD (Quantum Chromo
Dynamics) a gauge theory based on the symmetry group SU(3).
The group SU(3) could in matrix language be described as the set of unitary 
$3 \times 3$ matrices whose determinants equals unity. Such a matrix
has 8 independent parameters that corresponds to the eight gluons that
mediates the interaction. The quarks, which are the only particles to be 
affected by the strong interaction, carry a strong charge. This comes in three
different types, whimsically labeled R(ed) G(reen) and B(lue). The antiquarks 
carries the corresponding anticharges labeled  $\mrm{\bar{R}}$, 
$\mrm{\bar{G}}$ and $\mrm{\bar{B}}$. 

Quarks never exists freely, but are always combined into 
color singlet states called hadrons.
There are three fundamental ways to build a hadron: 
\begin{enumerate}
\item equal combinations of colors RGB
\item equal combinations of anticolors  
$\mrm{\overline{R}\,\overline{G}\,\overline{B}}$
\item equal combinations of a color and its anticolor 
$\mrm{R\overline{R}+G\overline{G}+B\overline{B}}$
\end{enumerate} 
No 1 corresponds to baryons, no 2 to antibaryons and no 3 to mesons and
antimesons. The quarks and gluons are collectively called partons,
this term stems from the days before the quark model was well established.
QCD is a non-Abelian field theory; this corresponds to the
fact that the gluons carry color charge themselves.
The gluons can thus interact with each other.   
This self-interaction give rise to the confinement of quarks into hadrons.
The field lines of the color field between two quarks forms 
a narrow tube when the quarks separation increases, instead of spreading
out as the electromagnetic field lines.
The energy in the color field is thus proportional to
the distance between the quarks. When the potential energy is big enough
a quark antiquark pair is created. The consequence is that there can
never exist a free color charged quark. Another consequence of the 
self-interactions is that the
equations of QCD can only be solved perturbatively for short distances,
i.e. high energies. For larger distances the perturbative approach fails.
Since the equations of QCD are too complex to be solved directly one must 
use some kind of simplified picture. One such approximation is the 
Lund model. 

This article is focused on the QCD treatment of the
$\mrm{e^{+}e^{-}}\rightarrow$ hadrons event.
The article is organized as follows: 
Section 2 presents the two standard methods used to describe the perturbative
phase of QCD, i.e. matrix elements and parton showers.
It also contains a description of fragmentation of quarks into hadrons and 
the jets that these hadrons form.
Section 3 describes the hybrid of matrix elements and parton showers, 
the main result of my diploma work. Section 4 shows the theoretical
predictions of the hybrid compared with those of the matrix elements and the 
parton showers.
%
%
\section{Existing descriptions of the $\mrm{e^{+}e^{-}}\rightarrow$ hadrons 
event}
\begin{figure}[t]
\begin{center}  
\begin{picture}(120,80)(0,0)
\Line(20,60)(40,40)
\Line(20,20)(40,40)
\Vertex(40,40){2}
\Photon(40,40)(80,40){3}{4}
\Vertex(80,40){2}
\Line(80,40)(100,60)
\Line(80,40)(100,20)
\Text(10,60)[]{$\mrm{e^{+}}$}
\Text(10,20)[]{$\mrm{e^{-}}$}
\Text(60,50)[]{$\mrm{\gamma / Z^{0}}$}
\Text(110,60)[]{q}
\Text(110,20)[]{$\mrm{\overline{q}}$}
\end{picture}
\end{center}
\caption{\label{fig:0-order}$\mrm{e^{+}e^{-}}\rightarrow \mrm{q\overline{q}}$}
\end{figure}
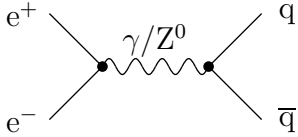

%
%
One experimental way to study QCD is the collision of $\mrm{e^{+}}$ with 
$\mrm{e^{-}}$.
This method has the advantage that the process is very clean in the sense 
that both $\mrm{e^{+}}$ and $\mrm{e^{-}}$ are considered to be fundamental 
point-like
particles. The process could to first order in $\as$ be described by the 
Feynman diagram in figure \ref{fig:0-order}.
The zeroth order cross section $\sigma_{0}$ is approximately given by
\begin{equation} 
\sigma_{0}\sim\left(
\frac{A}{p^{2}}+\frac{B~p^{2}}{(p^{2}-m^{2}_{\mrm{Z^{0}}})^{2}
+m^{2}_{\mrm{Z^{0}}}\Gamma_{Z^{0}}^{2}}
+\mrm{interference}\right)
\label{eqn:0-order}
\end{equation}
where $p$ is the four momentum of either $\mrm{\gamma}$ or $\mrm{Z^{0}}$.
The zeroth order cross section contains no QCD corrections except for a 
factor 3 counting the number of possible quark colors.
The first term corresponds to the $\mrm{\gamma}$ channel while the second term 
corresponds to the $\mrm{Z^{0}}$ channel. There is also a 
$\mrm{\gamma}$/$\mrm{Z^{0}}$ interference term between the two channels.
The factors A and B depends on the electro-magnetic and the weak 
coupling constants.
The quantity $p^{2}$ equals the total energy in the center of momentum frame
that coincides with the rest frame of the detector. The cross section
has two peaks at $m_{\mrm{\gamma}}=0$ GeV and $m_{\mrm{Z^{0}}}=91.2$ GeV.
Experiments designed for detailed studies of QCD are therefore 
preferably run at $E_{\mrm{cm}}=91.2$ GeV in order to obtain maximum 
statistics. 
%
%
\subsection{Matrix elements}
%
%
The perturbative solution to the QCD equations is expanded in terms of 
$\as$, the strong coupling constant. 
The matrix elements are the terms of this expansion. They can be represented by
Feynman diagrams. Each three-parton vertex involving the strong interaction is 
assigned a factor
$\sqrt{\as}$. The four-gluon vertex has a factor $\as$ but it does not appear 
until in 
higher orders than considered here.
Because the probability amplitudes are squared to obtain the 
cross section the number of strong interaction vertices equals the order in 
$\as$.
The strong coupling constant is, despite its name, 
a function of $Q^{2}=E_{\mrm{cm}}^{2}$ where cm refers to the center of
momentum frame of the whole event. 
To first order $\as$ is given by \cite{running-as}
\begin{equation}
\as (Q^{2}) = \frac{12\pi}{ (33-2n_{\mrm{f}})\log (Q^{2}/\Lambda^{2}) }~.
\label{as-Q}
\end{equation}
The integer $n_{\mrm{f}}$ is the number of quark flavors available at the 
given energy, typically 4 or 5. 
As seen from expression (\ref{as-Q}) $\as (Q^{2})$ decreases 
with increasing energies and therefore becomes small for short-distance
interactions. We say that QCD is asymptotically free.
The parameter $\Lambda$ is a free parameter that has to be determined by
experiments. According to experiments $\Lambda$ lies in the interval
0.2 to 0.3 GeV~. 

\begin{figure}[ht]
\begin{center}
\begin{picture}(140,100)(0,0)
\Line(20,10)(60,50)
\Line(20,90)(60,50)
\Vertex(60,50){2}
\Photon(60,50)(80,50){3}{2}
\Vertex(80,50){2}
\Line(80,50)(120,10)
\Line(80,50)(120,90)
\Vertex(100,30){2}
\Gluon(100,30)(120,50){3}{2}
\Text(10,90)[]{$\mrm{e^{+}}$}
\Text(10,10)[]{$\mrm{e^{-}}$}
\Text(130,90)[]{$\qbar$}
\Text(130,50)[]{g}
\Text(130,10)[]{q}
\end{picture}
\hspace{2cm}
\begin{picture}(140,100)(0,0)
\Line(20,10)(60,50)
\Line(20,90)(60,50)
\Vertex(60,50){2}
\Photon(60,50)(80,50){3}{2}
\Vertex(80,50){2}
\Line(80,50)(120,10)
\Line(80,50)(120,90)
\Vertex(100,70){2}
\Gluon(100,70)(120,50){3}{2}
\Text(10,90)[]{$\mrm{e^{+}}$}
\Text(10,10)[]{$\mrm{e^{-}}$}
\Text(130,90)[]{$\qbar$}
\Text(130,50)[]{g}
\Text(130,10)[]{q}
\end{picture}
\end{center}
\caption{\label{fig:1-order}$\mrm{e^{+}e^{-}\rightarrow qg\qbar}$}
\end{figure}
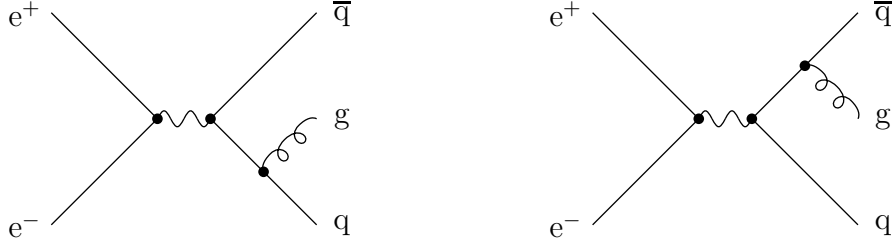

\begin{figure}[ht]
\begin{center}
\scalebox{0.8}{
\begin{picture}(140,100)(0,0)
\Line(20,10)(60,50)
\Line(20,90)(60,50)
\Vertex(60,50){2}
\Photon(60,50)(80,50){3}{2}
\Vertex(80,50){2}
\Line(80,50)(120,10)
\Line(80,50)(120,90)
\Gluon(100,70)(100,30){3}{3.5}
\Vertex(100,70){2}
\Vertex(100,30){2}
\Text(10,90)[]{$\mrm{e^{+}}$}
\Text(10,10)[]{$\mrm{e^{-}}$}
\Text(130,90)[]{$\qbar$}
\Text(130,10)[]{q}
\end{picture}
\hspace{1.5cm}
\begin{picture}(140,100)(0,0)
\Line(20,10)(60,50)
\Line(20,90)(60,50)
\Vertex(60,50){2}
\Photon(60,50)(80,50){3}{2}
\Vertex(80,50){2}
\Line(80,50)(120,10)
\Line(80,50)(120,90)
\GlueArc(100,30)(15,-45,135){3}{4}
\Vertex(89,41){2}
\Vertex(111,19){2}
\Text(10,90)[]{$\mrm{e^{+}}$}
\Text(10,10)[]{$\mrm{e^{-}}$}
\Text(130,90)[]{$\qbar$}
\Text(130,10)[]{q}
\end{picture}
\hspace{1.5cm}
\begin{picture}(140,100)(0,0)
\Line(20,10)(60,50)
\Line(20,90)(60,50)
\Vertex(60,50){2}
\Photon(60,50)(80,50){3}{2}
\Vertex(80,50){2}
\Line(80,50)(120,10)
\Line(80,50)(120,90)
\GlueArc(100,70)(15,-135,45){3}{4}
\Vertex(89,59){2}
\Vertex(111,81){2}
\Text(10,90)[]{$\mrm{e^{+}}$}
\Text(10,10)[]{$\mrm{e^{-}}$}
\Text(130,90)[]{$\qbar$}
\Text(130,10)[]{q}
\end{picture}}
\end{center}
\caption{\label{fig:virtual}first order virtual corrections}
\end{figure}
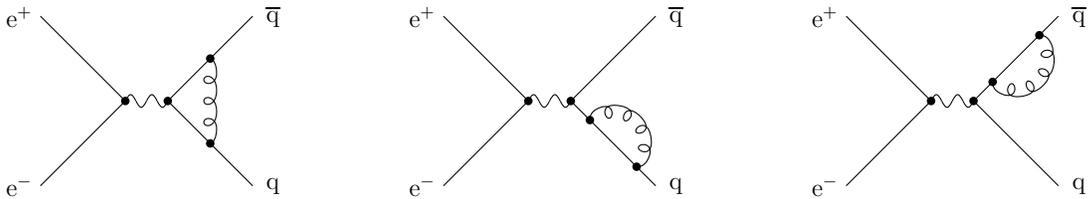

The matrix elements are in this article only showed
explicitly to first order in $\as$  because the second order matrix elements
\cite{ref:second-order} are far too complex.
The amplitude for a 3-parton event is to first order given by
the Feynman diagrams in figure \ref{fig:1-order}.
The cross section $\sigma_{1}$ is given explicitly as
\begin{equation}
\label{3p-CS}
\frac{1}{\sigma_{0}}\,\frac{\mrm{d}\sigma_{1}}
{\mrm{d}x_{1}\,\mrm{d}x_{2}}=
\frac{\as}{2\pi}\, \frac{4}{3}\, \frac{x_{1}^{2} + x_{2}^{2}}
{(1-x_{1})(1-x_{2})}
\hspace{1cm}
x_{i}=\frac{2E_{i}}{E_{\mrm{cm}}}~~~i=1,2,3
\end{equation}
where 1,2,3 corresponds to q,$\mrm{\overline{q}}$,g and  the energies 
$E_{i}$ are given in the center of momentum frame.
The masses of the partons are assumed to be 0. It is then easy to show that 
$0<x_{i}<1$. Expression (\ref{3p-CS}) is divergent in the region where at least
one of $x_{1}$ or $x_{2}$ approaches 1. The case where only one of them
approaches 1 is 
called the collinear singularity, because the other two parton momenta
become collinear. The other case when both approaches 1 is called the soft
singularity because the gluon energy becomes 0. 
The total first order cross section $\sigma_{\mrm{tot}}$ is given by
\begin{equation}
\sigma_{\mrm{tot}}=\left(1+\frac{\as}{\pi}\right)\sigma_{0}=
\sigma_{0}+\sigma_{1}+\sigma_{\mrm{virt}}
\end{equation}
where $\sigma_{\mrm{virt}}$ is the first order virtual corrections created by 
the interference between the diagrams in figure \ref{fig:0-order} and 
\ref{fig:virtual}.
The above mentioned divergences in $\sigma_{1}$ are cancelled by divergences in 
$\sigma_{\mrm{virt}}$ thus rendering $\sigma_{\mrm{tot}}$ finite.
The divergences in $\sigma_{1}$ is therefore no real analytical problem but the
use of Monte Carlo methods requires $0\leq\sigma_{1}/\sigma_{\mrm{tot}}\leq 1$.
To handle this we introduce cuts in the phase space by 
demanding that $y_{jk}>y_{\mrm{min}}$~. The variable $y_{jk}$ is given by 
$y_{jk}=(p_{j}+p_{j})^{2}/E^{2}_{\mrm{cm}}=1-x_{i}$  
where indices i,j,k are all different.

By plotting the expression
\begin{equation}
\frac{\sigma_{1}}{\sigma_{0}}=\int_{y_{ij}>y_{\mrm{min}}}\frac{1}{\sigma_{0}}\,
\frac{\mrm{d}\sigma_{1}}{\mrm{d}x_{1}\,\mrm{d}x_{2}}\,\mrm{d}x_{1}\,
\mrm{d}x_{2}=
\frac{\as}{2\pi}\, \frac{4}{3}\, \int_{y_{ij}>y_{\mrm{min}}} 
\frac{(1-y_{23})^{2} + (1-y_{13})^{2}}{y_{23}\,y_{13}}\,\mrm{d}y_{23}\,
\mrm{d}y_{13}
\end{equation}
as a function of $y_{\mrm{min}}$ with $\as\simeq 0.15$ one notices that 
$y_{\mrm{min}}$ must be at least $0.01$ to ensure that 
$\sigma_{1}/\sigma_{\mrm{tot}}\leq1$~. 
This corresponds to a minimum mass
\begin{equation} 
m_{\mrm{min}}=\sqrt{y_{\mrm{min}}}E_{\mrm{cm}}~. 
\label{eqn:m-min}
\end{equation}
At the $Z^{0}$ resonance we obtain $m_{\mrm{min}}\simeq 9$ GeV.
The running of $\as$ sets a limit for the perturbative approach at 
$m_{ij}\gtrsim 1$ GeV. The result is that the gluons in the region 1 to 9 GeV  
are missed. As seen from expression (\ref{eqn:m-min}) the region of discarded
gluons increases as a linear function of the energy $E_{\mrm{cm}}$.
This is a severe shortcoming of the matrix element approach.

The amplitudes for a 4-parton event to second order in $\as$
is given by the Feynman diagrams listed in fig \ref{4P-hist}.
\begin{figure}[htbp]
\begin{center}
\scalebox{0.7}{
\begin{picture}(120,150)(0,20)
\Text(0,170)[]{1}
\Photon(0,100)(20,100){3}{2}
\Vertex(20,100){2}
\Line(20,100)(100,160)
\Line(20,100)(100,40)
\Vertex(60,130){2} 
\Vertex(60,70){2}
\Gluon(60,130)(100,115){4}{3}
\Gluon(60,70)(100,85){4}{3}
\Text(110,160)[]{q}
\Text(110,40)[]{$\mrm{\overline{q}}$}
\Text(110,115)[]{g}
\Text(110,85)[]{g}
\end{picture}
\hspace{2cm}
\begin{picture}(120,150)(0,20)
\Text(0,170)[]{2}
\Photon(0,100)(20,100){3}{2}
\Vertex(20,100){2}
\Line(20,100)(100,160)
\Line(20,100)(100,40)
\Vertex(47,80){2}
\Gluon(47,80)(100,120){4}{6}
\Vertex(73,60){2}
\Gluon(73,60)(100,85){4}{3}
\Text(110,160)[]{q}
\Text(110,40)[]{$\mrm{\overline{q}}$}
\Text(110,120)[]{g}
\Text(110,85)[]{g}
\end{picture}
\hspace{2cm}
\begin{picture}(120,150)(0,20)
\Text(0,170)[]{3}
\put(0,200){\rotatebox{180}{\reflectbox{
\Photon(0,100)(20,100){3}{2}
\Vertex(20,100){2}
\Line(20,100)(100,160)
\Line(20,100)(100,40)
\Vertex(47,80){2}
\Gluon(47,80)(100,120){4}{6}
\Vertex(73,60){2}
\Gluon(73,60)(100,85){4}{3}}}}
\Text(110,160)[]{q}
\Text(110,40)[]{$\mrm{\overline{q}}$}
\Text(110,120)[]{g}
\Text(110,85)[]{g}
\end{picture}} 
\end{center}
\begin{center}
\scalebox{0.7}{
\begin{picture}(120,150)(0,20)
\Text(0,170)[]{4}
\put(0,200){\rotatebox{180}{\reflectbox{
\Photon(0,100)(20,100){3}{2}
\Vertex(20,100){2}
\Line(20,100)(100,160)
\Line(20,100)(100,40)
\Vertex(47,80){2}
\Gluon(47,80)(73,100){4}{2}
\Vertex(73,100){2.5}
\Gluon(73,100)(100,120){4}{2.5}
\Gluon(73,100)(100,80){4}{2.5}}}}
\Text(110,160)[]{q}
\Text(110,40)[]{$\mrm{\overline{q}}$}
\Text(110,120)[]{g}
\Text(110,85)[]{g}
\end{picture} 
\hspace{2cm}
\begin{picture}(120,150)(0,20)
\Text(0,170)[]{5}
\Photon(0,100)(20,100){3}{2}
\Vertex(20,100){2}
\Line(20,100)(100,160)
\Line(20,100)(100,40)
\Vertex(47,80){2}
\Gluon(47,80)(73,100){4}{2}
\Vertex(73,100){2.5}
\Gluon(73,100)(100,120){4}{2.5}
\Gluon(73,100)(100,80){4}{2.5}
\Text(110,160)[]{q}
\Text(110,40)[]{$\mrm{\overline{q}}$}
\Text(110,120)[]{g}
\Text(110,80)[]{g}
\end{picture}
\hspace{2cm}
\begin{picture}(120,150)(0,20)
\end{picture}}
\end{center} 
\begin{center}
\scalebox{0.7}{
\begin{picture}(120,150)(0,20)
\Text(0,170)[]{6}
\put(0,200){\rotatebox{180}{\reflectbox{
\Photon(0,100)(20,100){3}{2}
\Vertex(20,100){2}
\Line(20,100)(100,160)
\Line(20,100)(100,40)
\Vertex(47,80){2}
\Gluon(47,80)(73,100){4}{2}
\Vertex(73,100){2.5}
\Line(73,100)(100,120)
\Line(73,100)(100,80)}}}
\Text(110,160)[]{q}
\Text(110,40)[]{$\mrm{\overline{q}}$}
\Text(110,120)[]{q}
\Text(110,80)[]{$\mrm{\overline{q}}$}
\end{picture}
\hspace{2cm}
\begin{picture}(120,150)(0,20)
\Text(0,170)[]{7}
\Photon(0,100)(20,100){3}{2}
\Vertex(20,100){2}
\Line(20,100)(100,160)
\Line(20,100)(100,40)
\Vertex(47,80){2}
\Gluon(47,80)(73,100){4}{2}
\Vertex(73,100){2.5}
\Line(73,100)(100,120)
\Line(73,100)(100,80)
\Text(110,160)[]{q}
\Text(110,40)[]{$\mrm{\overline{q}}$}
\Text(110,120)[]{q}
\Text(110,80)[]{$\mrm{\overline{q}}$}
\end{picture}
\hspace{2cm}
\begin{picture}(120,150)(0,20)
\end{picture}}
\end{center}
\captive{4-parton histories to second order in $\as$
\label{4P-hist}}
\end{figure}
The diagrams can be divided into three categories: double-bremsstrahlung (1-3),
triple-gluon vertex (4-5) and secondary q\={q}-production (6-7).
The cross sections for reaction
$\mrm{e^{+}e^{-} \rightarrow \gamma^{*} / Z^{0} 
\rightarrow q \overline{q}gg}$ and reaction $\mrm{e^{+}e^{-} \rightarrow 
\gamma^{*} / Z^{0} \rightarrow q \overline{q} q \overline{q}}$
are given by \cite{ref:4pcs}
\begin{eqnarray}
\sigma_{\mrm{q\overline{q}gg}}(y_{ij}) & = &
\sigma_{0}
\left[C_{\mrm{F}}^{2}A(y_{ij})+\left(C_{\mrm{F}}^{2}-\frac{1}{2}C_{\mrm{F}}
C_{\mrm{A}}\right)B(y_{ij})
+ C_{\mrm{F}}C_{\mrm{A}}C(y_{ij})\right] 
\label{eqn:c-section1} \\
\sigma_{\mrm{q\overline{q}q\overline{q}}}(y_{ij}) & = &
\sigma_{0}
\left[C_{\mrm{F}}T_{\mrm{F}}n_{\mrm{f}}D(y_{ij})+\left(C_{\mrm{F}}^{2}
-\frac{1}{2}C_{\mrm{F}}C_{\mrm{A}}\right)E(y_{ij})
\right]
\label{eqn:c-section2}
\end{eqnarray}
where $n_{\mrm{f}}$ is the number of active quark flavors and 
$y_{ij}=m_{ij}^{2}/E_{\mrm{cm}}^{2}$ is the normalized two-body invariant
mass with indices i and j running over the four partons.
The functions $A(y_{ij})\ldots E(y_{ij})$ are group independent and contains
the full kinematical dependence of the cross section.
A set of five independent $y_{ij}$ are needed to specify the kinematical
configuration of a 4-parton event.
The so-called Casimir factors $C_{\mrm{F}}$, $C_{\mrm{A}}$, $T_{\mrm{F}}$ 
are respectively a measure of the coupling strengths of the reactions
$\mrm{q}\rightarrow \mrm{qg}$, $\mrm{g}\rightarrow \mrm{gg}$ and 
$\mrm{g}\rightarrow \mrm{q\overline{q}}$.
Thus the first term in expression (\ref{eqn:c-section1}) correspond to diagrams
1 to 3 in figure \ref{4P-hist} and the third term correspond to diagrams
4 and 5. The second term in expressions (\ref{eqn:c-section1}) and 
(\ref{eqn:c-section2}) are formed by the quantum mechanical interference.
The Casimir factors $C_{\mrm{F}}$, $C_{\mrm{A}}$, $T_{\mrm{F}}$ are given 
directly by the symmetry group. For QCD the group is SU(3) and the Casimir
factors have the values $C_{\mrm{F}}=4/3$, $C_{\mrm{A}}=3$ and 
$T_{\mrm{F}}=1/2$. 
%

%
%
\subsection{Parton showers}
%
%

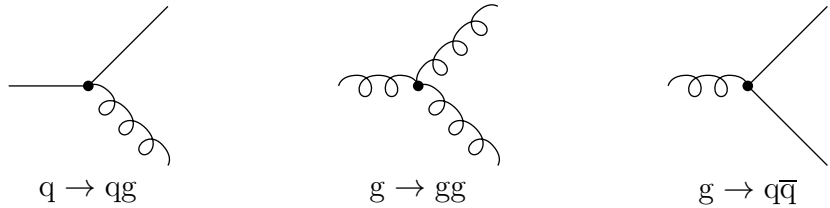
\begin{figure}[htbp]
\begin{center}
\begin{picture}(60,80)(0,0)
\Line(0,50)(30,50)
\Vertex(30,50){2}
\Line(30,50)(60,80)
\Gluon(30,50)(60,20){4}{3}
\Text(30,10)[]{$\mrm{q}\rightarrow \mrm{qg}$}
\end{picture}
\hspace{2cm}
\begin{picture}(60,80)(0,0)
\Gluon(0,50)(30,50){4}{2}
\Vertex(30,50){2}
\Gluon(30,50)(60,80){4}{3}
\Gluon(30,50)(60,20){4}{3}
\Text(30,10)[]{$\mrm{g}\rightarrow \mrm{gg}$}
\end{picture}
\hspace{2cm}
\begin{picture}(60,80)(0,0)
\Gluon(0,50)(30,50){4}{2.5}
\Vertex(30,50){2}
\Line(30,50)(60,80)
\Line(30,50)(60,20)
\Text(30,10)[]{$\mrm{g}\rightarrow \mrm{q\overline{q}}$}
\end{picture}
\end{center}
\caption{\label{fig:branchings}The basic branchings in a parton shower}
\end{figure}

A parton shower is a kind of semi-classical approximation of parton events in
the sense that every parton in the event has a given history with a specified
four-momentum. Under the simulation each particle is allowed to branch into
two new particles. The three basic branchings are shown in figure
\ref{fig:branchings}. 
By iterating these basic branches a final state with an arbitrary number
of partons may be constructed.
The differential probabilities for each branching are given by the 
so-called \mbox{Altarelli-Parisi} equations of evolution 
\cite{ref:Altarelli-Parisi}:
\begin{equation}
\mrm{d}P_{a\rightarrow bc} = \frac{ \as (Q^{2}) }{ 2\pi }\,
\frac{\mrm{d}m^{2}_{a} }{ m^{2}_{a} }\, P_{a\rightarrow bc}(z)\,\mrm{d}z~.
\label{Alt-Paris}
\end{equation}
The splitting kernels $P_{a\rightarrow bc}(z)$ are given by the following 
expressions depending on the type of branching:
\begin{eqnarray} 
P_{\mrm{q}\rightarrow \mrm{qg}}(z) & = & \frac{4}{3}\,\frac{1+z^{2}}{1-z}~,
\nonumber\\
P_{\mrm{g}\rightarrow \mrm{gg}}(z) & = & 3\,\frac{(1-z(1-z))^{2}}{z(1-z)}~,
\nonumber\\  
P_{\mrm{g}\rightarrow \mrm{q\overline{q}}}(z) & = & \frac{\mrm{n_{f}}}{2}\,
(z^{2}+(1-z)^{2})~.
\label{z-kernels}
\end{eqnarray}
The splitting kernels can be obtained from the 3-parton matrix elements.
By defining $z=x_{1}$ expression (\ref{3p-CS})
turns into:
\begin{eqnarray}
\nonumber
\label{ME-PS}
\mrm{d} P_{\mrm{q} \rightarrow \mrm{q g}} & = & \frac{ \mrm{d}\sigma }
{ \sigma_{0} } =
\frac{\as}{2\pi}\, \frac{4}{3}\, \frac{x_{1}^{2} + x_{2}^{2}}
{(1-x_{1})(1-x_{2})}\, \mrm{d}x_{1}\,\mrm{d}x_{2} \\
\nonumber
& = & \frac{\as}{2\pi}\, \frac{4}{3}\, \frac{z^{2} + (1-y_{13})^{2}}
{(1-z)y_{13}}\, \mrm{d}z\,\mrm{d}y_{13} \\
\nonumber
& \simeq & \frac{\as}{2\pi}\, \frac{4}{3}\, \frac{\mrm{d}
m^{2}_{\mrm{q}}}{m^{2}_{\mrm{q}}} 
\,\frac{z^{2} + 1} {1-z}\, \mrm{d}z \\
& = & \frac{\as}{2\pi}\, \frac{\mrm{d}m^{2}_{\mrm{q}}}{m^{2}_{\mrm{q}}}\, 
P_{\mrm{q} \rightarrow \mrm{q g}}(z)\, \mrm{d}z~. 
\end{eqnarray}
In the last step comparison with expression (\ref{Alt-Paris}) gives the 
expression for $P_{\mrm{q} \rightarrow \mrm{q g}}(z)$.
The expressions $P_{\mrm{g} \rightarrow \mrm{g g}} $ and 
$P_{\mrm{g} \rightarrow \mrm{q \bar{q}}}$ could be obtained in a similar way.
As seen from the approximation in expression (\ref{ME-PS}) the parton shower
only gives a good description in the limit $y_{13}\rightarrow 0$, i.e. when
the partons produced in the branching becomes collinear.
Probabilities larger than unity in the collinear region correspond to 
possibility of a quark emitting more than one gluon.
Parton showers are therefore complementary to the matrix elements 
in the sense that they are well behaved below the  $y_{\mrm{min}}$ cut at 
9 GeV.
The exact definition of $z$ and $Q^{2}$ varies amongst different algorithms but
generally $z$ is some variant of energy sharing between the two daughters and
$Q^{2}$ is a function of $m^{2}_{a}$ and $z$. Furthermore
the expression $\mrm{d}m^{2}_{\mrm{q}}/m^{2}_{\mrm{q}}$ in equation 
(\ref{Alt-Paris}) can be replaced by $\mrm{d}(m^{2}_{\mrm{q}}f(z))/
(m^{2}_{\mrm{q}}f(z))$ where $f(z)$ is 
an arbitrary well-behaved function of $z$.

The probability that a parton does not branch between an initial mass $m$
and a minimum value $m_{\mrm{min}}$ is given by the integration and 
exponentiation of expression (\ref{Alt-Paris}):
\begin{equation}
S_{a}(m)=\exp \left \{ -\int^{m^{2}}_{m^{2}_{\mrm{min}}} 
\frac{\mrm{d}m'^{2}}{m'^{2}}
\int^{z_{\mrm{max}}(m')}_{z_{\mrm{min}}(m')} 
\frac{ \alpha_{\mrm{s}}(Q^{2}) }{ 2\pi }\, \mrm{d}z\, P_{a\rightarrow bc}(z) 
\right \}
\label{Sudakov}
\end{equation}
where $m_{\mrm{min}}$ is a small cutoff mass that is used to regularize
the collinear divergences in $z$ and the infrared divergences in $m^{2}$.
Expression (\ref{Sudakov}) is called the Sudakov form factor.
An analogue to expression (\ref{Sudakov}) is the formula
\begin{displaymath}
\frac{N}{N_{0}}=e^{ -\lambda t }
\end{displaymath}
which gives the fraction of radioactive nucleus still remaining after a 
time $t$. The probability that a parton with a maximum mass $m_{\mrm{max}}$
will branch in the interval $[m^{2},m^{2} + dm^{2}]$ is given by 
\begin{equation}
P_{a}(m_{\mrm{max}}^{2},m^{2}) = S_{a}(m_{\mrm{max}}^{2}) \frac{\mrm{d}}
{\mrm{d}m^{2}}
\left\{ \frac{1}{ S_{a}(m^{2}) } \right\} \mrm{d}m^{2}~.
\end{equation}
A branching with a specific $m$ is then selected using a Monte Carlo method.
A random number $R$ is chosen uniformly in the interval $(0,1)$ and the
equation 
\begin{equation}
S_{a}(m^{2}) = \frac{S_{a}(m_{\mrm{max}}^{2})}{R} \label{m-equation}
\end{equation}
is solved for m. Unfortunately is it often impossible to solve 
equation (\ref{m-equation}) analytically but appropriate
numerical methods exists. Whenever the lower limit $m=m_{\mrm{min}}$ is
reached the particle is put on its mass-shell and the evolution of the 
particle is terminated.

An independent evolution of each parton in the shower overestimates
the total amount of evolution. The coherence effect can to some extent be
included by imposing angular ordering, i.e. by requiring that the angle 
between two daughter
partons decreases as one goes from the initial to the final partons.
A consequence of the angular ordering is a slower multiplicity growth of
partons and a depletion of parton production at small energies.

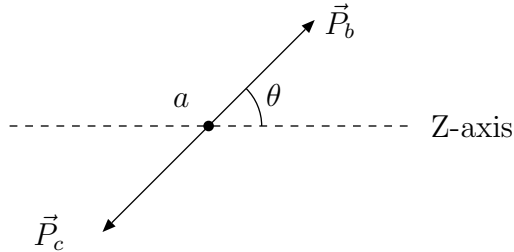
\begin{figure}[ht]
\begin{center}
\begin{picture}(200,120)(0,0)
\DashLine(0,60)(150,60){3}
\Vertex(75,60){2}
\CArc(75,60)(20,0,45)
\LongArrow(75,60)(115,100)
\LongArrow(75,60)(35,20)
\Text(125,100)[]{$\vec{P}_{b}$}
\Text(15,20)[]{$\vec{P}_{c}$}
\Text(100,72)[]{$\theta$}
\Text(175,60)[]{Z-axis}
\Text(65,70)[]{$a$}
\end{picture}
\end{center}
\caption{\label{fig:z-def}The branching $a\rightarrow bc$ in the rest frame of 
$a$.}
\end{figure}

%
%
%
%
As previously mentioned the definition of $Q^{2}$ and $z$ 
in the parton shower varies between
the different parton shower algorithms. In {\sc JETSET} \cite{ref:jetset},
the event generator that I've been using, the argument $Q^{2}$ 
in $\as$ is given by
\begin{equation}
Q^{2} = z(1-z)m^{2}_{a}\sim p_{T}^{2}
\end{equation}
where $p_{T}$ is the component of the daughters momentum transverse 
to the momentum of the parent. The definition of $Q^{2}$ is motivated by the
attempt to include coherence effects. 
To choose a definition for $z$ we first look at the system in the rest-frame of 
parton $a$ and with the z-axis collinear to the velocity of $a$ in the parton 
shower CM-frame, see figure \ref{fig:z-def}.
If we now translate to the  parton showers CM-frame
it is straightforward to show that
\begin{equation}
z = \frac{E_{b}}{E_{a}} = \frac{1+\beta\mrm{\cos}\theta}{2} 
\hspace{2cm} 
\beta = \frac{(p_{a})_{z}}{E_{a}} \label{z-theta}
\end{equation}
under the assumption that $b$ and $c$ are massless.
In the infinite-momentum frame, $\beta\rightarrow 1$, this reduces to
$z=(1+\cos\theta)/2$.
If we now decide that relation (\ref{z-theta}) should hold even if
$b$ and $c$ are massive then, after some calculations, $z$ turns
out to be given by:
\begin{eqnarray} 
z &=& \frac{ m^{2}_{a} }{ \lambda }\frac{ E_{b} }{ E_{a} } - 
\frac{m^{2}_{a} - \lambda + m^{2}_{b} - m^{2}_{c}}{2\lambda}
\label{z-def}
\\
\mrm{with~~}
\lambda &=& \sqrt{(m^{2}_{a} - m^{2}_{b} - m^{2}_{c})^{2} - 4m^{2}_{b}\,
m^{2}_{c}}~.
\nonumber
\end{eqnarray}
%
%
%
\subsection{Fragmentation}
%
%
At large times, when the partons produced by a matrix element or a 
parton shower become more separated, the strong
coupling constant $\as$ increases to the
limit where perturbative calculations become impossible.
Because of the increasing strength of the strong interaction all
color-charged partons are forced to combine into colorless states,
the so-called hadrons. It is this process of free quarks and gluons 
turning into hadrons that is called fragmentation.
The process is not yet understood in the context of the fundamental
QCD equations, so some kind of phenomenological method is needed.
These methods can be divided into 3 categories: 
independent fragmentation, string fragmentation and cluster fragmentation.
One example of string fragmentation is the `Lund model'.
To explain the string model we first look at how it works in the
simplest case with a q$\qbar$ color-singlet. Calculations with 
lattice QCD indicate that the energy in the color-field is proportional
to the distance between the partons. This leads to the picture of
a color flux tube that is stretched between the two particles.
The flux tube is uniform along its length and has a transverse size
of roughly 1 fm. When the distance between the endpoints q and $\qbar$
increases the energy in the tube becomes large enough to create two new
partons $\mrm{q'}\qbar'$. The result is two new color-singlets 
q$\qbar'$ and $\mrm{q'}\qbar$ each with its own flux tube. 
The process is then repeated until only on-mass-shell hadrons remains. 

In the more complex situation with the initial configuration 
qg$\qbar$, shown in figure \ref{fig:string},  
one draws a tube from the quark to the gluon and then to the anti-quark.
\begin{figure}[ht]
\begin{center}
\begin{picture}(100,100)(0,0)
\LongArrow(50,30)(10,0)
\LongArrow(50,30)(90,0)
\Gluon(50,30)(50,70){3}{3}
\DashLine(10,0)(50,70){4}
\DashLine(90,0)(50,70){4}
\Vertex(50,70){1}
\Text(0,0)[]{q}
\Text(100,0)[]{$\qbar$}
\Text(50,80)[]{g}
\end{picture}
\end{center}
\caption{\label{fig:string}String drawing for a $\mrm{qg\overline{q}}$ event}
\end{figure}
The gluon must be placed inside the string because the gluon has a 
color-charge given by a color and a anti-color. The charges of the string
q -- g -- $\qbar$ is of the form R -- $\mrm{\overline{R}}$G -- 
$\mrm{\overline{G}}$. The sub-strings q -- g 
and g -- $\qbar$ are then evolving as in the q$\qbar$ example except for one
meson that is produced at the gluon-corner consisting of one parton from 
each sub-string. When the partonic state contains $2n$ quarks
then $n$ strings are formed, each containing a quark, an antiquark and an
arbitrary number of gluons. 

The procedure above only produces mesons.
One easy way to obtain the production of baryons is to treat a diquark
in a color antitriplet state as an ordinary antiquark.
The strings can then either break by quark-antiquark or antidiquark-diquark
pair production. An alternative model is the `popcorn' model in which
the antidiquark-diquark pair production is replaced by two consecutive
quark-antiquark pair productions. The `popcorn' model gives a less strong 
correlation between the color and momentum space of the baryon and
the antibaryon coming from the same pair production.

Many of the particles produced in the fragmentation process are unstable 
particles with very short lifetimes. 
The simulation of the decay process is governed by tables over the possible 
decay modes and
their corresponding branching ratios. Normally it is assumed that 
decay products are distributed according to phase space, i.e. there is no
dynamics involved in their relative distribution.
 
%
%
\subsection{Jets and their angular distributions}
%
%
The angular distribution of the particles detected in an $e^{+}e^{-}$ collision 
is not uniform. Instead the particles are clustered together in so-called
jets. The directions of the jets are strongly correlated to the
directions of the hardest partons in the perturbative phase. 
Since there is a continuum of differently separated parton emissions there
is no unique jet definition.
In order to determine the number of jets and their corresponding four-momenta 
some kind of cluster algorithm has to be used on the final particles.
One type of cluster algorithm are those based on binary joining.
In these algorithms all particles are considered to be 
separate clusters at the start.
Then a distance measure $d_{ij}$ is calculated for every pair of clusters.
If the smallest $d_{ij}$ is smaller than some given $d_{\mrm{join}}$ 
then the corresponding clusters are joined into a new cluster.
The procedure is then repeated for the remaining clusters until all clusters
are separated by a distance greater than $d_{\mrm{join}}$.
The final clusters are then the seeked-for jets.
One possible definition of $d_{ij}$ used in the  
{\sc JADE} scheme \cite{ref:jade1} is
%
\begin{equation}
d_{ij}=\frac{2E_{i}E_{j}(1-\cos\theta_{ij})}{E^{2}_{\mrm{vis}}}~.
\label{eqn:jade1}
\end{equation}
Here $E_{\mrm{vis}}$ is the total visible energy of the event.
In an event simulation and in an ideal experiment $E_{\mrm{vis}}$ equals
$E_{\mrm{cm}}$. The dimensionless nature of $d_{ij}$ in expression 
(\ref{eqn:jade1}) makes it suitable for comparing results at different
c.m. energies.
When given the jets from a 4-parton event there is no unique pairing of
partons with jets. The quark-jets are harder than the gluon-jets
on the average, but the difference smears out for the individual
jets under the fragmentation and the clustering process.  
With the four jets enumerated in decreasing energy order 
the first two jets preferentially correspond to the initial q\={q}-par
while the two last jets preferentially correspond to the emitted gluons
or the secondary produced q\={q}-par. 
The uncertainty in the parton-jet pairing prevents one from accessing the full
five-dimensional distribution of the 4-parton event.
Instead there is need for simpler measures that are less sensitive to the above
discussed uncertainties.
The following angular distributions for 4-jet events are examples of such 
measures:
\begin{enumerate}
\item The Bengtsson-Zerwas correlation \cite{ref:BZangle}:
\begin{equation}
\label{BZ}
\mrm{\cos} \chi_{\mrm{BZ}} =  \frac{|
(\vec{p}_{1} \times \vec{p}_{2}) \cdot
(\vec{p}_{3} \times \vec{p}_{4})| 
}{
| \vec{p}_{1} \times \vec{p}_{2} | 
| \vec{p}_{3} \times \vec{p}_{4} |
} ~,
\end{equation}
\item The modified Nachtmann-Reiter variable \cite{ref:NRangle}:
\begin{equation}
\label{NR}
| \mrm{\cos} \theta^{\ast}_{\mrm{NR}} | =  \frac{|
(\vec{p}_{1} - \vec{p}_{2}) \cdot
(\vec{p}_{3} - \vec{p}_{4})|  
}{
| \vec{p}_{1} - \vec{p}_{2} | 
| \vec{p}_{3} - \vec{p}_{4} |
} ~,
\end{equation}
\item The cosine of the angle between the two jets with lowest energy
\cite{ref:alpha}:
\begin{equation}
\label{alpha34}
\mrm{\cos} \alpha_{34} =  \frac{
\vec{p}_{3} \cdot \vec{p}_{4}
}{
| \vec{p}_{3} || \vec{p}_{4} |
}~.
\end{equation}
\end{enumerate}
The three variables are not orthogonal but still sufficiently different to
provide complementary information. 
The variable $\mrm{\cos}\chi_{\mrm{BZ}}$ 
measure the angle between the plane spanned by the two most energetic jets
and the plane spanned by the two least energetic jets.
The $\mrm{g}\rightarrow \mrm{gg}$ vertex in QCD tends to align the two planes,  
contrary to the $\mrm{g}\rightarrow\mrm{q}\qbar$ vertex that favors a 
perpendicular relation between the planes. This is due to the different
spins of the quark and the gluon.
The variable could therefore be used to measure the difference between QCD
and an Abelian variant of QCD that lacks the $\mrm{g}\rightarrow\mrm{gg}$ 
vertex.
The Abelian variant is not a serious competitor to QCD because it doesn't 
explain the running of $\as$, it is rather a counter-example used to prove
the non-Abelian character of QCD. 
The original Nachtmann-Reiter variable measure the angle
$\theta_{\mrm{NR}}$ between the two vectors $\vec{p}_{1}$ and $\vec{p}_{3}$. 
The $\mrm{g}\rightarrow \mrm{gg}$ vertex tends to decrease the 
$\theta_{\mrm{NR}}$ angle contrary to the $\mrm{g}\rightarrow\mrm{q}\qbar$ that
tends to increase $\theta_{\mrm{NR}}$. This is again due to the different spins
of the quark and the gluon.
The modified angle $\theta^{\ast}_{\mrm{NR}}$ becomes equal to 
$\theta_{\mrm{NR}}$ in the limit when parton 1 and 2 are back-to-back whereas 
when the angles between parton 1 and 2 or between parton 3 and 4 decrease,
$\theta^{\ast}_{\mrm{NR}}$ is roughly equal to $\chi_{\mrm{BZ}}$.
The observable $\theta^{\ast}_{\mrm{NR}}$ is thus complementary 
to $\chi_{\mrm{BZ}}$. It is used in order to distinguish between
two-gluon final states and secondary q$\mrm{\overline{q}}$-production.
The variable $\mrm{\cos}\alpha_{34}$ is used in order to distinguish 
between the double-bremsstrahlung and the triple-gluon vertex.

%
\section{Creating a forced parton shower}
%
%
The second-order QCD matrix elements give a very good agreement 
with experimental data on the angular distributions of the four-jet
events in $\mrm{e}^{+}\mrm{e}^{-}$ collisions 
at the $\mrm{Z^{0}}$ resonance \cite{ref:ME-good}.
Unfortunately the description of the subjet structure is quite poor. 
The alternative approach, parton showers, gives a good description of the 
subjet structure but is worse than matrix elements when it comes to the angular
distributions. To enhance the theoretical accuracy there are different paths 
to follow. One way is to calculate matrix elements to a higher order. 
Unfortunately this is impossible in practice because of the enormous amount of
work needed. Another way is to first use the matrix element and then apply a
parton shower on the resulting partons. The problem here is that
the parton shower is normally formulated so that it starts from two initial 
partons in order get
all its internal procedures right. For example the angular ordering procedure
doesn't have any initial angle to start from.
In this report the attempt is to combine the matrix
element and the parton shower in a more subtle way. 
The combination is done in the sense
that a parton-shower is forced to follow a history derived from a 
matrix-element
in the first steps of its evolution. The forced parton shower has been 
implemented as a set of
{\sc Fortran77} subroutines which in turn heavily rely on the 
event generator {\sc JETSET 7.410}.
The whole procedure is described in detail in the following text:
\begin{enumerate}

\item
The 4-parton matrix element procedure based on the paper 
\cite{ref:second-order} is executed for given values on $E_{\mrm{cm}}$
and $y_{\mrm{min}}$. 
The result is the four-momentum and the 
particle codes of the final 
particles in either reaction $\mrm{e^{+}e^{-} \rightarrow \gamma^{*} / Z^{0} 
\rightarrow q \overline{q}gg}$ or in reaction $\mrm{e^{+}e^{-} \rightarrow 
\gamma^{*} / Z^{0} \rightarrow q \overline{q} q \overline{q}}$.
How these final particles were formed is not 
known; this is due to the underlying quantum mechanical theory that
treats the process as a superposition of all possible histories.

\item
The Feynman diagrams corresponding to all 4 jet histories of second order
in $\as$ are shown in fig \ref{4P-hist}. The enumeration of the diagrams is 
the same as
is used in the program source code to label the different histories.   
The particle codes are examined in order to decide if the event, generated
in step 1, was a 
$\mrm{q\overline{q}gg}$, event number 1--5, or a 
$\mrm{q\overline{q}q\bar{q}}$ 
event, number 6--7. When the event type has been narrowed down to one of these
categories it still remains to select one of the relevant histories. 
The exact choice of history can not be uniquely determined, due to the quantum
mechanical interference, but a probability can be assigned
to each history. The probability for a diagram is given by the product of
the branching probabilities for the two vertices. The branching probabilities  
are given by expression (\ref{Alt-Paris}) but with the approximation 
(necessary for the matrix-element approach)
that $\as(Q^{2})$ is assumed to be equal for the two vertices.
The phase space is chosen as $\mrm{d}m^{2}\mrm{d}z$ in order to have a
consistency between the parton-shower algorithm and the matrix-element
descriptions, cf. eq. (\ref{ME-PS}). 
For example the relative probability for the history shown in figure 
\ref{fig:trace-hist} is given by:
\begin{equation}
P=P_{1\rightarrow 34}P_{4\rightarrow 56}=\frac{1}{m_{1}^{2}}\frac{4}{3}
\frac{1+z^{2}_{34}}{1-z_{34}}\cdot\frac{1}{m_{4}^{2}}
3\frac{(1-z_{56}(1-z_{56}))^{2}}{z_{56}(1-z_{56})}
\end{equation}
where 
\begin{eqnarray}
m_{1}^{2} &=& p_{1}^{2} = (p_{3}+p_{5}+p_{6})^{2}~, \\
m_{4}^{2} &=& p_{4}^{2} = (p_{5}+p_{6})^{2}~, \nonumber
\end{eqnarray}
and z values are evaluated using expression (\ref{z-def}).
\begin{figure}[htbp]
\begin{center}
\begin{picture}(120,140)(0,30)
\Text(5,110)[]{0}
\Text(30,122)[]{1}
\Text(30,80)[]{2}
\Text(55,140)[]{3}
\Text(55,97)[]{4}
\Text(85,125)[]{5}
\Text(85,75)[]{6}
\Text(110,160)[]{q}
\Text(110,40)[]{$\mrm{\overline{q}}$}
\Text(110,120)[]{g}
\Text(110,85)[]{g}
\put(0,200){\rotatebox{180}{\reflectbox{
\Photon(0,100)(20,100){3}{2}
\Vertex(20,100){2}
\Line(20,100)(100,160)
\Line(20,100)(100,40)
\Vertex(47,80){2}
\Gluon(47,80)(73,100){4}{2}
\Vertex(73,100){2.5}
\Gluon(73,100)(100,120){4}{2.5}
\Gluon(73,100)(100,80){4}{2.5}}}}
\end{picture} 
\end{center}
\caption{\label{fig:trace-hist} 4-parton history no 4}  
\end{figure}
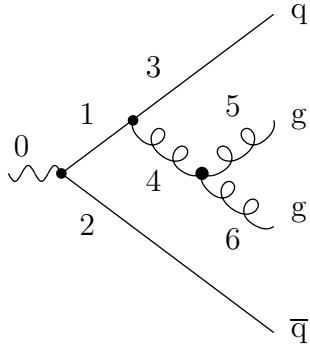
Once the probabilities have been calculated a specific history is chosen
in one of the following ways:
\begin{itemize}
\item
Monte Carlo: a history is chosen at random but according to the 
relative probabilities calculated above.
\item
Winner takes all: the history with the largest probability is chosen.
\end{itemize}
\item
Once a history is chosen the values of $z$, $m_{a}^{2}$ and 
a $\varphi$-angle is recorded for each of the two vertices.  
The two $\varphi$-angles are obtained in the following way.
First the coordinates are rotated so that the initial quark momentum
is in the positive z direction, then a second rotation is done so that the 
the mother of the considered vertex has its momentum in the positive z 
direction. These rotations consist of a rotation in the $xy$-plane followed
by a rotation in the $xz$-plane. The need to specify
a specific rotation procedure is due to the fact that the orientation of a 
rigid body is determined by three Euler angles. 
The $\varphi$-angle is then given by the polar representation
of one of the daughters momentum:
\begin{eqnarray}
p_{x} & = & r\sin\theta\cos\varphi \\
p_{y} & = & r\sin\theta\sin\varphi \nonumber \\
p_{z} & = & r\cos\theta \nonumber
\end{eqnarray}
In a q$\rightarrow$qg or a $\qbar\rightarrow\qbar$g vertex the daughter 
is chosen to be the quark.

\item 
A modified version of the parton shower procedure in {\sc JETSET}
is executed. Given the chosen history the procedure forces the $z$-value
and the $m_{a}^{2}$-value of the two vertices in the corresponding diagram 
to be as close as possible to the values given by the matrix element.
The forced $z$ value and the $z$ value given by the matrix element differs 
slightly, see figure \ref{fig:z-diff}. 
\begin{figure}[htbp]
\begin{center}
\rotatebox{90}{
\scalebox{1.7}{\mbox{
\begin{picture}(0,0)(0,0)
\put(-130,10){$\mrm{d}N/\mrm{d}~(z_{\mrm{final}}-z_{\mrm{forced}})$}
\end{picture}}}}
\scalebox{0.5}{\rotatebox{-90}{
\mbox{\epsfig{file=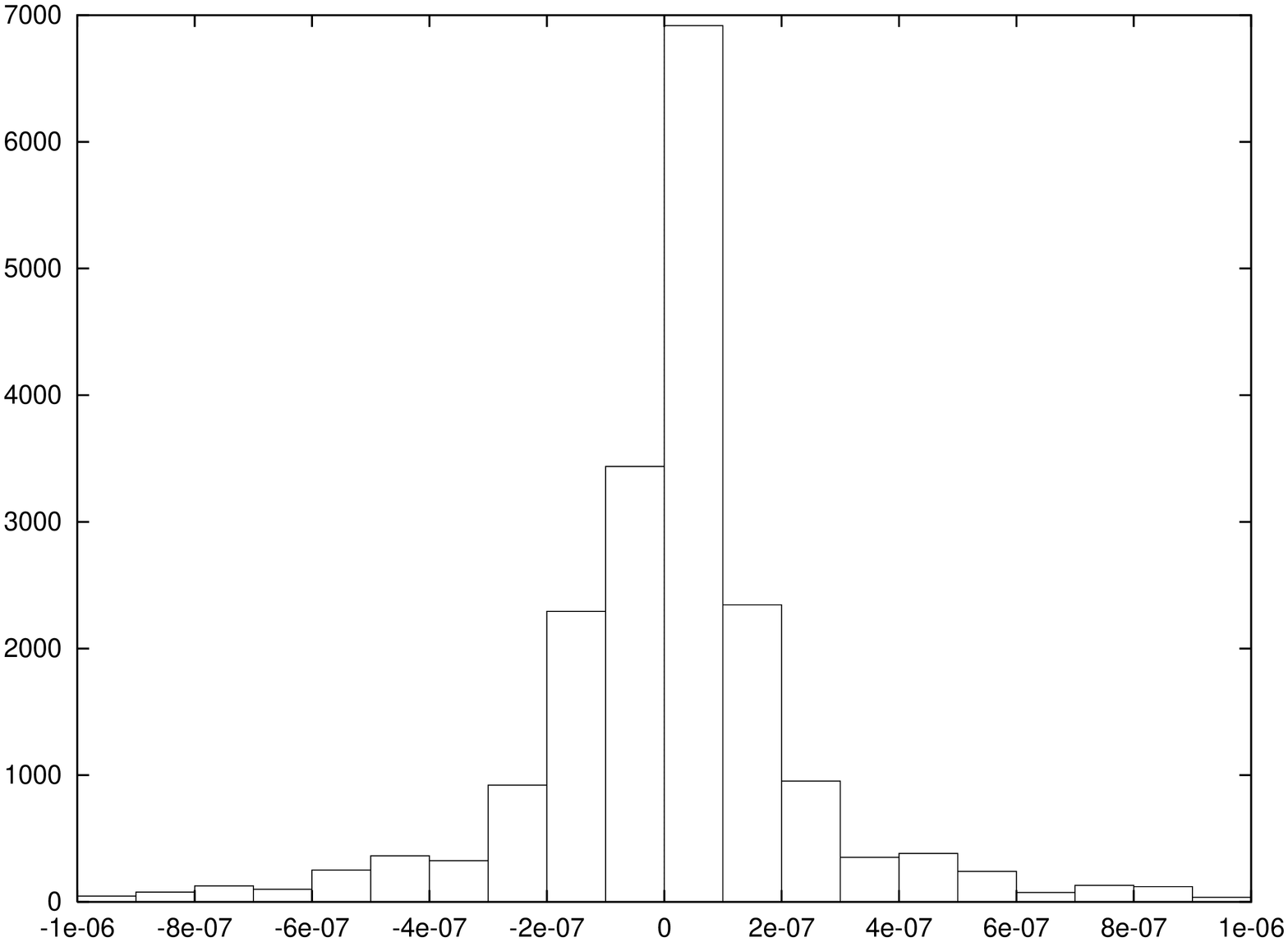}}}}
\\
\scalebox{1.7}{\mbox{
\begin{picture}(20,15)(0,0)
\put(0,0){$z_{\mrm{final}}-z_{\mrm{forced}}$}
\end{picture}}} 
\end{center}
\captive{\label{fig:z-diff} distribution of 
$z_{\mrm{final}}-z_{\mrm{forced}}$~.}
\end{figure}
This is due to the fact that the final partons given by the matrix 
element is on the mass-shell while the corresponding partons in the parton 
shower might be virtual particles who branches into new particles. 
The difference is fortunately small, roughly $10^{-6}$, compared with the
upper limit $z_{max}\simeq 1$.
Thus the kinematics is slightly different in the two cases.
The difference in kinematics is illustrated by figure \ref{fig:angle-diff}
that shows the distribution of the angle between the parton given by 
the matrix elements and the corresponding forced parton.
The angular differences are somewhat larger than the $z$ differences but
they are still small. The larger angular differences is mainly explained 
by the fact that the angle between the daughters in a branching decreases 
when the daughters obtain non-zero masses.
\begin{figure}[htbp]
\begin{center}
\rotatebox{90}{
\scalebox{1.7}{\mbox{
\begin{picture}(0,0)(0,0)
\put(-110,0){$\mrm{d}N/\mrm{d}~\mrm{angle}$}
\end{picture}}}}
\scalebox{0.5}{\rotatebox{-90}{
\mbox{\epsfig{file=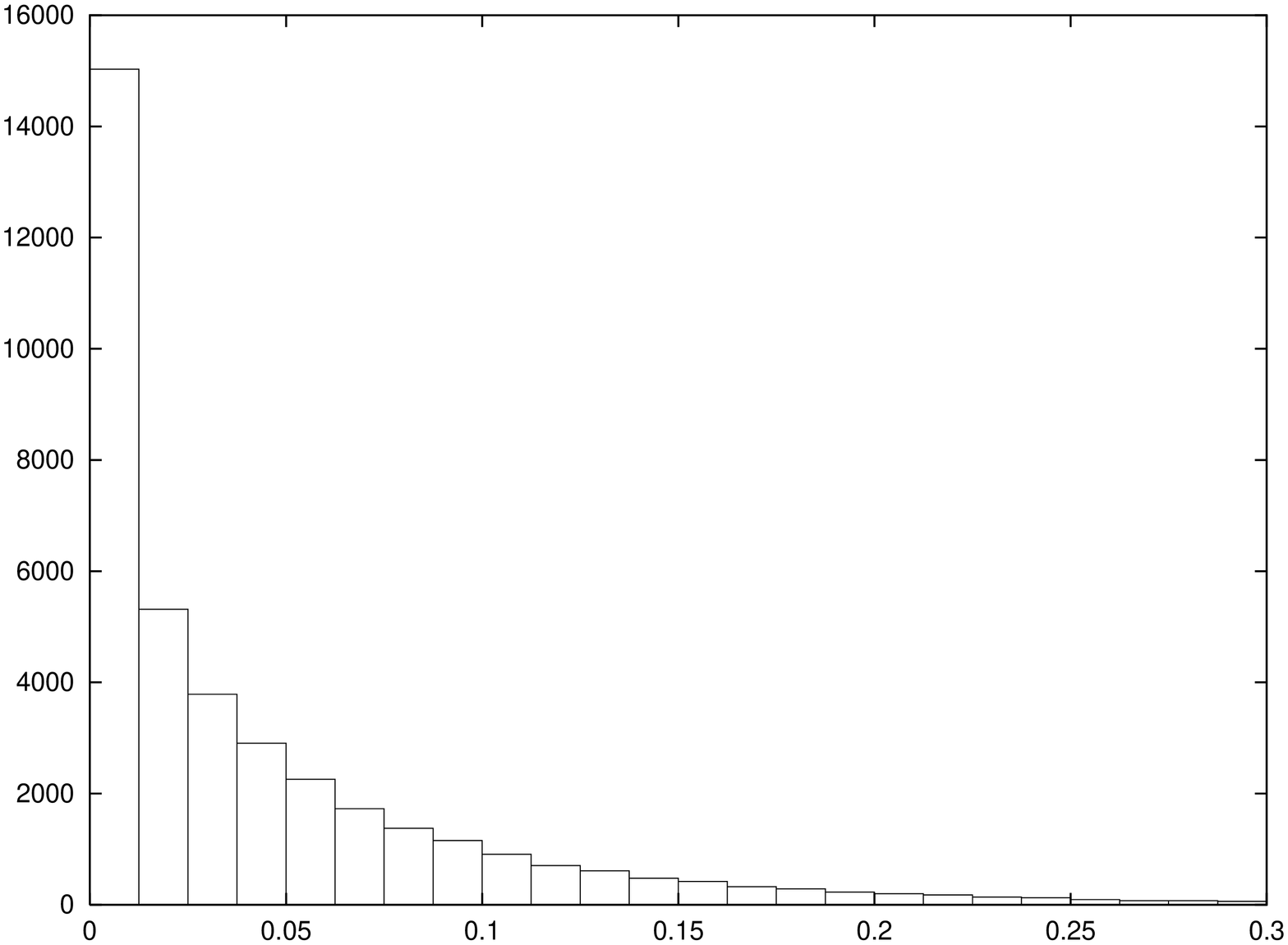}}}}
\\
\scalebox{1.7}{\mbox{
\begin{picture}(20,15)(0,0)
\put(-20,0){angle in radians}
\end{picture}}} 
\end{center}
\captive{\label{fig:angle-diff} Distribution of the 
angle between the forced parton and the corresponding matrix-elements 
parton.}
\end{figure} 
%
%

All other branchings are treated as before except that the masses of
the partons not included in the history are forced to have
masses below a given threshold. 
The matrix elements generate no emission below the mass cut at 9 GeV. 
The hybrid mass threshold must therefore be at least 9 GeV in order to
account for the emission missing from the matrix-element part. 
Since no 5-parton emission is 
generated by the second order matrix elements, one could allow a threshold
higher than 9 GeV in order to account for this potential emission.
However, if any such mass is larger than one of the forced masses, the
result would be a different history than the chosen one.
Thus one possible mass threshold is to choose it as the smallest of
the two forced masses. Another choice is a fixed mass threshold
at 13 GeV, a value that gives the same average multiplicity as the
original parton shower. This may be viewed as a pragmatical compromise
between the two extremes above.

To show the exact execution the forced parton-shower algorithm is traced on the
diagram shown in figure \ref{fig:trace-hist}.
\begin{description}

\item[a] 
The initial partons 1 and 2 are stored in the event array. Their 
four-momenta are given in the center of momentum system and their spatial
momentum is rotated so it coincides with the directions of the initial partons
given by the matrix element and the chosen history. The directions are in this
case given by $\vec{p}_{3}+\vec{p}_{5}+\vec{p}_{6}$ and $\vec{p}_{2}$. 
The partons are on the mass-shell. 

\item[b]
The mass of parton 1 and the $z$ value of $1\rightarrow 34$ are forced.
The mass of parton 2 is chosen by solving equation (\ref{m-equation}),
with the extra condition that it is below the previously mentioned 
threshold.
The type of branching for parton 2 is also decided. Because the parton is a 
antiquark the only available branching is 
$\qbar\rightarrow\qbar$g. If the branching parton is a gluon then one 
has to choose between 
branch g$\rightarrow$gg or branch g$\rightarrow$q$\qbar$. Once
the branch type is chosen the z value is chosen according to one of the 
probabilities in (\ref{z-kernels}).

\item[c]
The mass and z value of parton 3 is chosen in the same way as for parton 2
while the mass and z value of parton 4 is forced.
Then the $\varphi$ angle of parton 3 and 4 is forced around the axis of
parton 1.

\item[d]
Later on, after the masses of 5 and 6 have been selected,
the $\varphi$ angle of parton 5 and 6 is forced.
Then the parton shower is allowed to evolve in its usual way until all the
partons are left on their mass-shells.

\end{description}
\end{enumerate}
%
%
\section{Results}
\begin{figure}[htbp]
\begin{center}
\rotatebox{90}{
\scalebox{1.7}{\mbox{
\begin{picture}(0,0)(0,0)
\put(-130,0){$(1/N)\mrm{d}N/\mrm{d}~\cos\chi_{\mrm{BZ}}$}
\end{picture}}}}
\scalebox{0.5}{\rotatebox{-90}{
\mbox{\epsfig{file=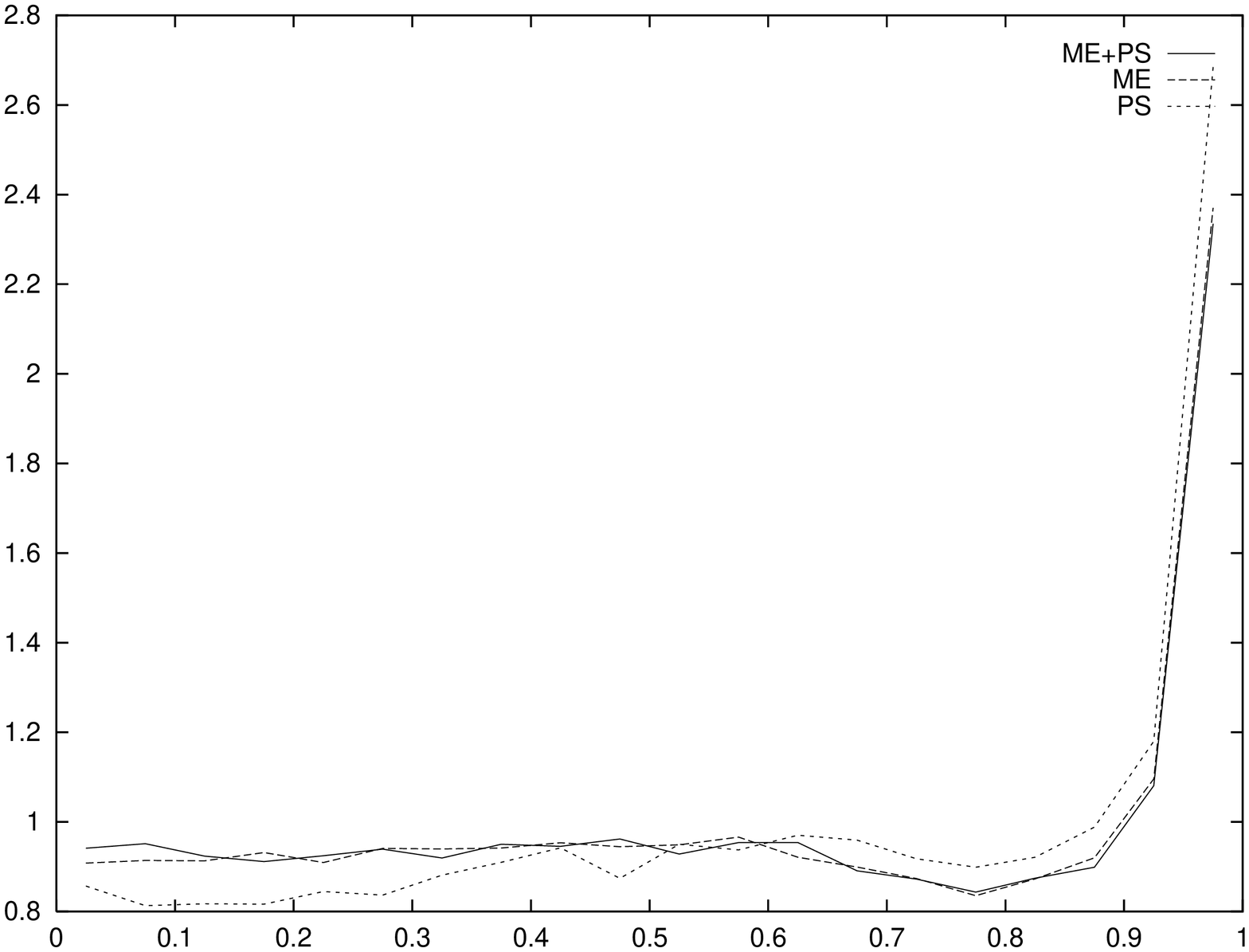}}}}
\\
\scalebox{1.7}{\mbox{
\begin{picture}(30,15)(0,0)
\put(0,0){$\cos\chi_{\mrm{BZ}}$}
\end{picture}}} 
\end{center}
\captive{\label{fig:BZ} $\mrm{\cos}\chi_{\mrm{BZ}}$ distributions}
\end{figure}

\begin{figure}[htbp]
\begin{center}
\rotatebox{90}{
\scalebox{1.7}{\mbox{
\begin{picture}(0,0)(0,0)
\put(-130,0){$(1/N)\mrm{d}N/\mrm{d}~|\mrm{\cos}\theta^{\ast}_{\mrm{NR}}|$}
\end{picture}}}}
\scalebox{0.5}{\rotatebox{-90}{
\mbox{\epsfig{file=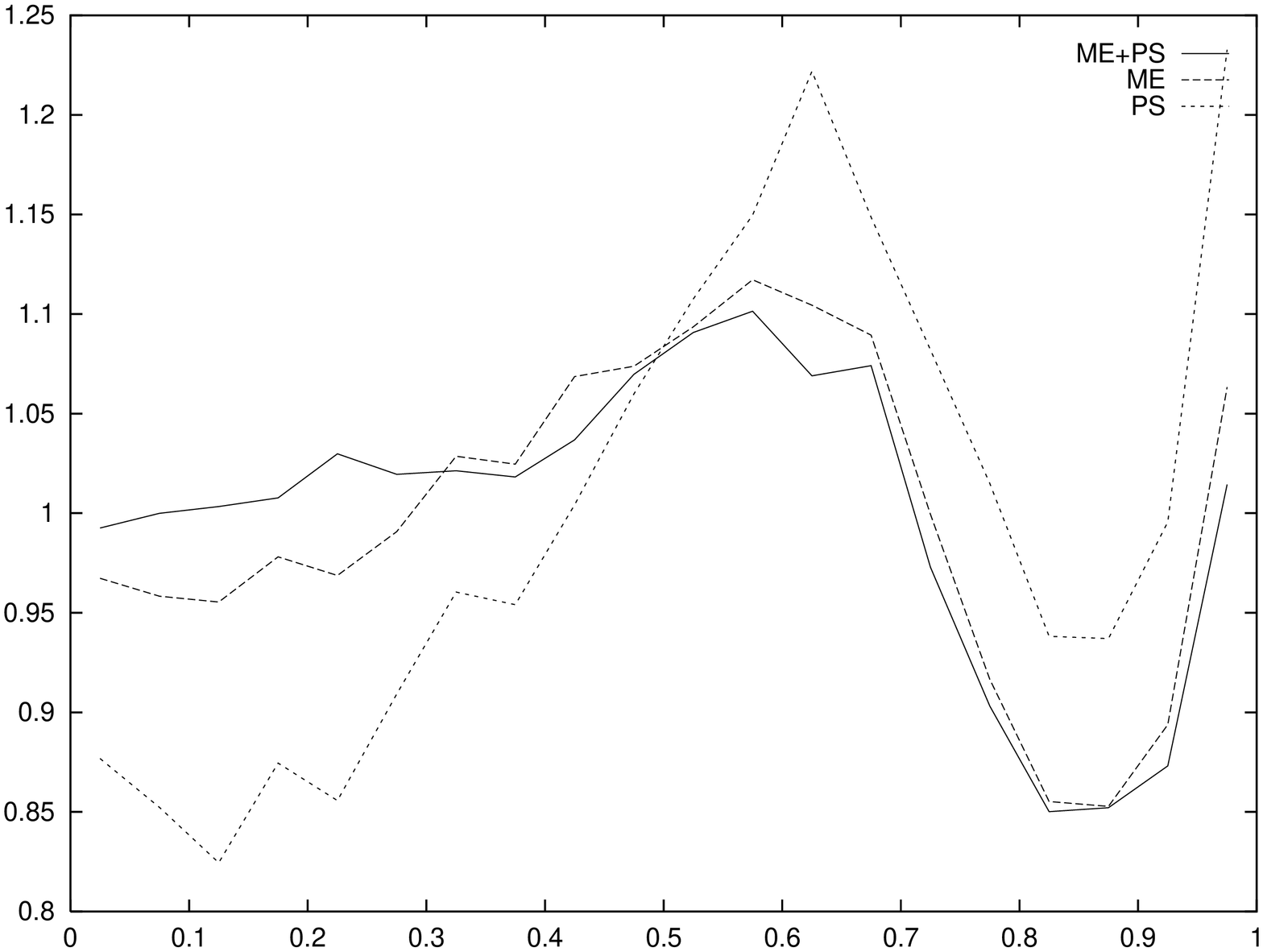}}}} 
\\
\scalebox{1.7}{\mbox{
\begin{picture}(40,15)(0,0)
\put(0,0){$|\cos\theta^{\ast}_{\mrm{NR}}|$}
\end{picture}}}
\end{center}
\captive{\label{fig:NR} $|\mrm{\cos}\theta^{\ast}_{\mrm{NR}}|$ distributions}
\end{figure}

\begin{figure}[htbp]
\begin{center}
\rotatebox{90}{
\scalebox{1.7}{\mbox{
\begin{picture}(0,0)(0,0)
\put(-130,0){$(1/N)\mrm{d}N/\mrm{d}~\mrm{\cos}\alpha_{34}$}
\end{picture}}}}
\scalebox{0.5}{\rotatebox{-90}{
\mbox{\epsfig{file=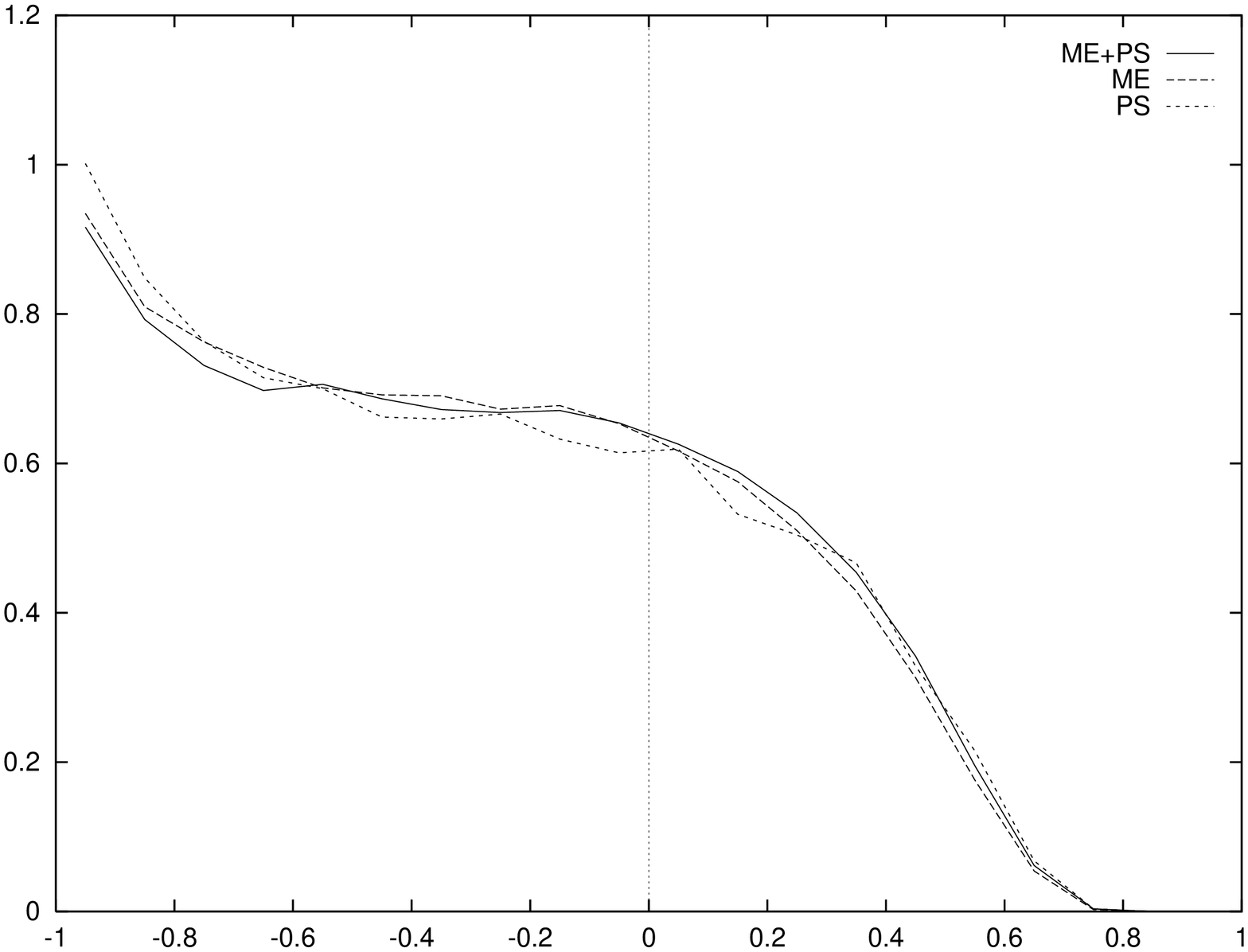}}}}
\\
\scalebox{1.7}{\mbox{
\begin{picture}(20,15)(0,0)
\put(0,0){$\cos\alpha_{34}$}
\end{picture}}} 
\end{center}
\captive{\label{fig:alpha34} $\mrm{\cos}\alpha_{34}$ distributions}
\end{figure}

\begin{figure}[htbp]
\begin{center}
\rotatebox{90}{
\scalebox{1.2}{\mbox{
\begin{picture}(0,0)(0,0)
\put(-190,0){$(1/N)\mrm{d}N/\mrm{d}~|\mrm{\cos}\theta^{\ast}_{\mrm{NR}}|/
|\mrm{\cos}\theta^{\ast}_{\mrm{NR}}|_{\mrm{ME}}$}
\end{picture}}}}
\scalebox{0.5}{\rotatebox{-90}{
\mbox{\epsfig{file=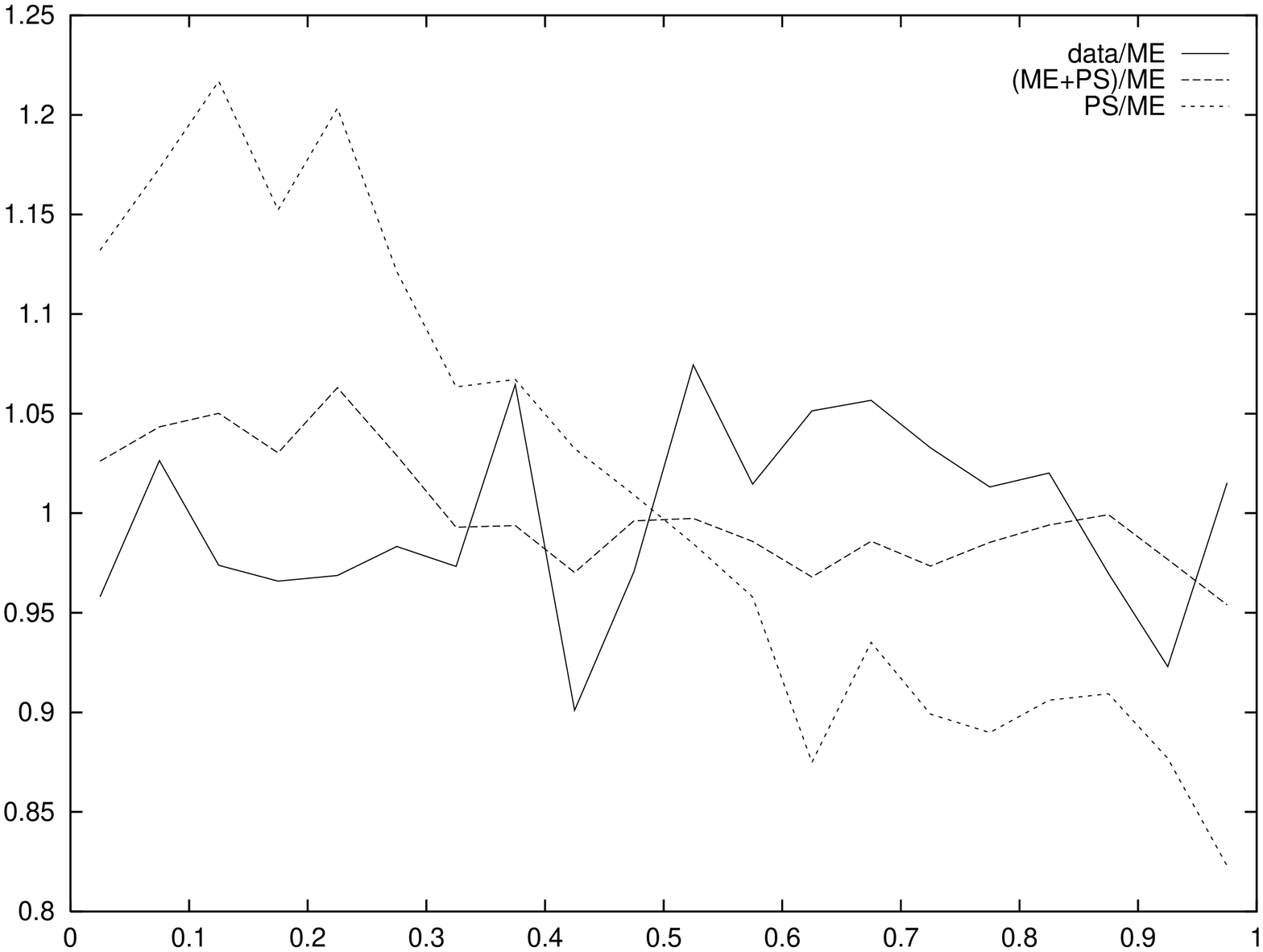}}}}
\\
\scalebox{1.7}{\mbox{
\begin{picture}(20,15)(0,0)
\put(-30,0){$|\mrm{\cos}\theta^{\ast}_{\mrm{NR}}|/
|\mrm{\cos}\theta^{\ast}_{\mrm{NR}}|_{\mrm{ME}}$}
\end{picture}}} 
\end{center}
\captive{\label{fig:kvoter}$|\mrm{\cos}\theta^{\ast}_{\mrm{NR}}|$ distributions
scaled by the ME distribution}
\end{figure}

%
%
In order to evaluate the hybrid between matrix elements and parton 
showers complete events are generated
with parton showers, matrix elements and the hybrid.
All events are generated at the $Z^{0}$-resonance 91.2 GeV.
As jet-finding algorithm the {\sc JADE} scheme (technically the {\sc P0} 
variant \cite{ref:jadeP0}) is used. Only four-jet events are kept for further 
analysis. 
The three different methods have been compared in the following
ways:
\begin{enumerate}

\item
The multiplicity distributions of charged  
particles are shown in figure \ref{fig:mult}.
The hybrid is executed with three different mass thresholds: fixed at 13 GeV 
(ME+PS I), fixed at 9 GeV (ME+PS II) and minimum of the two forced masses
(ME+PS III).
The lower limit on jet separation is set to $d_{\mrm{join}}=0.02$~.
This choice of $d_{\mrm{join}}$ give well separated jets and reasonable high
event statistics. The distribution of the parton shower (PS) and the three
hybrid distributions nearly coincide. 
These four distributions are also in good agreement with the experimental data
\cite{ref:multdata} from the DELPHI detector at the LEP collider. 
The only distribution that is apparently different and doesn't agree with the
experimental data is the matrix elements distribution (ME).

\begin{figure}
\begin{center}
\rotatebox{90}{
\scalebox{1.7}{\mbox{
\begin{picture}(0,0)(0,0)
\put(-130,0){$(1/N)\mrm{d}N/\mrm{d~(mult)}$}
\end{picture}}}}
\scalebox{0.5}{\rotatebox{-90}{
\mbox{\epsfig{file=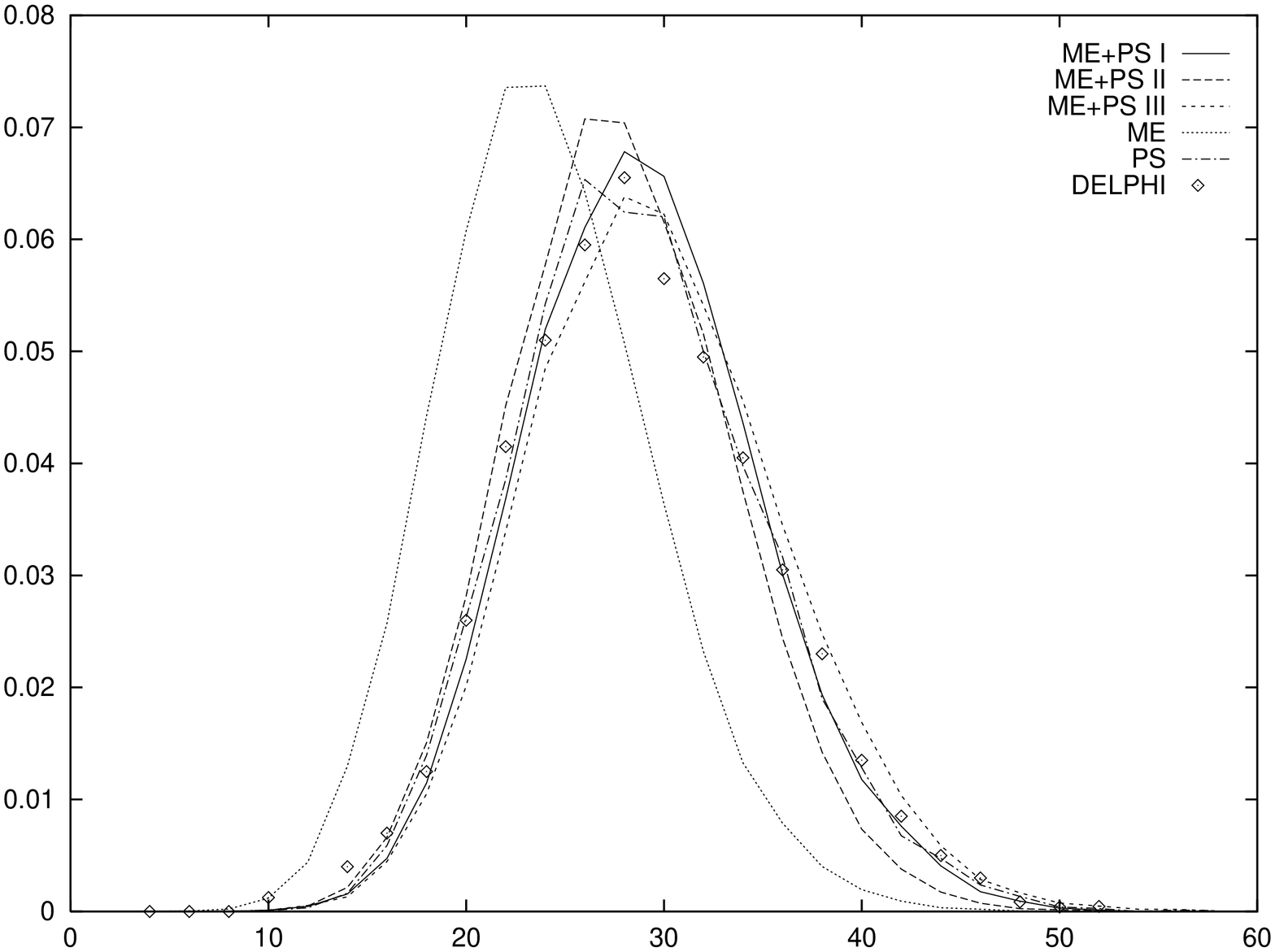}}}} 
\\
\scalebox{1.7}{\mbox{
\begin{picture}(50,15)(0,0)
\put(0,0){multiplicity}
\end{picture}}}
\end{center}
\captive{\label{fig:mult} Distributions of charged particle multiplicity}
\end{figure}

\item
The subjet multiplicity is plotted in figure \ref{fig:subjet-mult}.
The number of jets is plotted as a function of the lower limit on the jet 
separation $d_{\mrm{join}}\leq 0.03$ for events with four jets at 
$d_{\mrm{join}}=0.03$~.
Thus the subjet multiplicity gives a picture of how narrow the particles
are in the four original jets.
The events have been generated using the fixed mass threshold at 13 GeV.
The curve corresponding to the matrix elements has the strongest tendency 
towards 4 jets.
This is due to the fact that the matrix elements only generates 4 partons;
further jets can only appear by fluctuations in the fragmentation stage.

\begin{figure}
\begin{center}
\rotatebox{90}{
\scalebox{1.7}{\mbox{
\begin{picture}(0,0)(0,0)
\put(-120,-5){number of jets}
\end{picture}}}}
\scalebox{0.5}{\rotatebox{-90}{
\mbox{\epsfig{file=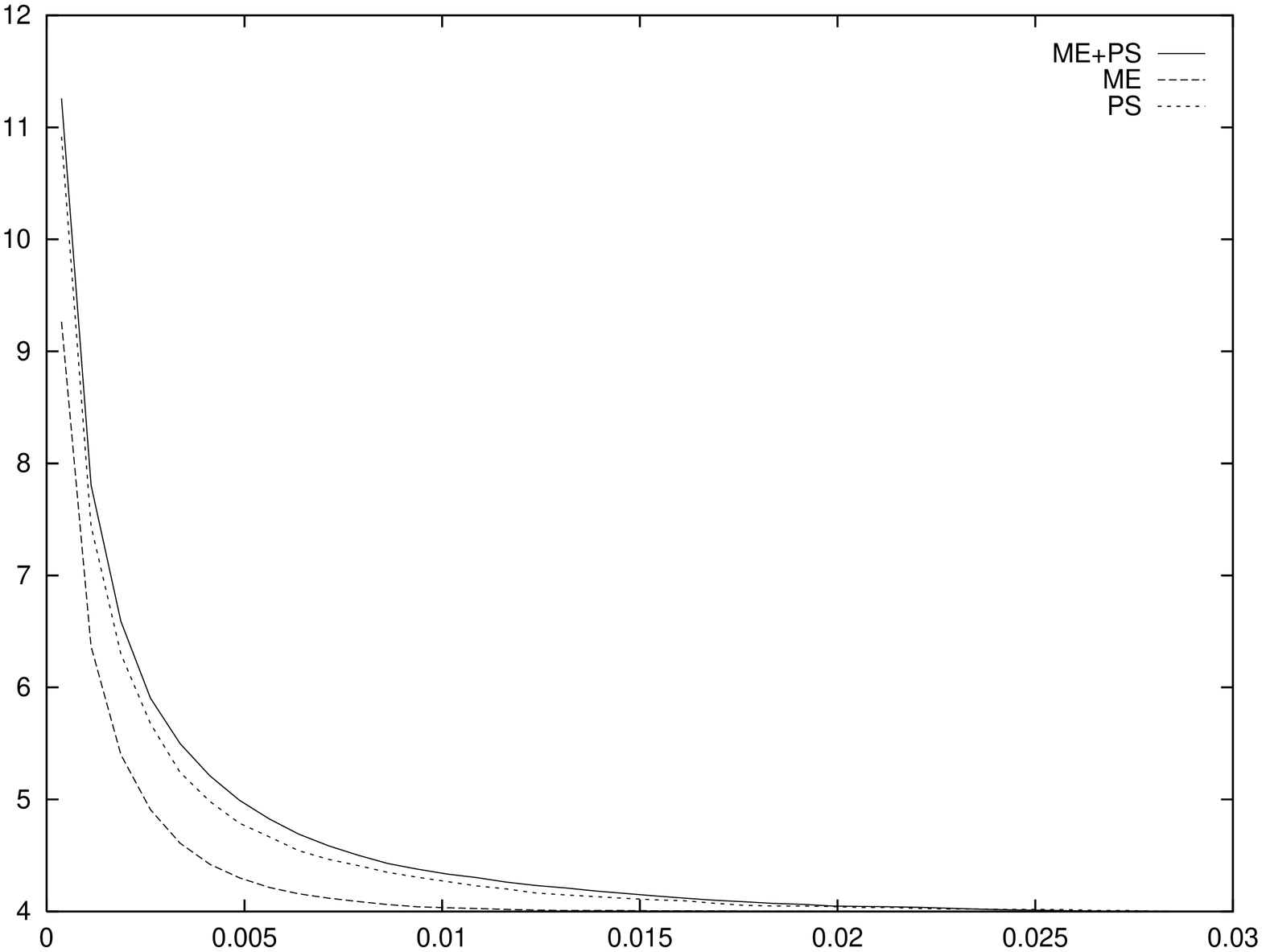}}}}
\\
\scalebox{1.7}{\mbox{
\begin{picture}(120,15)(0,0)
\put(0,0){minimal jetseparation $d_{\mrm{join}}$}
\end{picture}}} 
\end{center}
\captive{\label{fig:subjet-mult} subjet multiplicities}
\end{figure}

\item
The angular distributions defined in equations (\ref{BZ}-\ref{alpha34})
are plotted in figures \ref{fig:BZ} to \ref{fig:alpha34}.
The events have been generated using the fixed mass threshold at 13 GeV and
with the lower limit on jet separation set to $d_{\mrm{join}}=0.03$~.
The distributions of the matrix element (ME) and the hybrid 
(ME+PS) nearly coincides for all the three angular variables
while the distribution of the parton shower (PS) clearly deviates.
Comparison with data is complicated, since it is difficult exactly to
reproduce the experimental selection criteria,
but we know \cite{ref:ME-good} that the 
matrix elements describe them well. Thus  
$|\mrm{\cos}\theta^{\ast}_{\mrm{NR}}|$ distributions given by the hybrid,
the parton shower and by experimental data \cite{ref:ME-good} from the OPAL 
detector at LEP are scaled by the distribution given by the matrix elements.
The experimental data is scaled by the ME distribution given in the referred
paper while the other two distributions are scaled by the ME distribution
shown in figure \ref{fig:NR}.
The result is shown in figure \ref{fig:kvoter}.
The hybrid distribution clearly matches the experimental distribution better 
than the parton-shower distribution does. 
\end{enumerate}

\section{Summary and Outlook}
The results above shows that the hybrid simulation of hadronic
$\mrm{e^{+}e^{-}}$ events gives almost the same results as the
matrix elements in the case of the three angular variables
$\mrm{\cos} \chi_{\mrm{BZ}}$, $|\mrm{\cos}\theta^{\ast}_{\mrm{NR}}|$
and $\mrm{\cos}\alpha_{34}$. The hybrid also matches the parton showers
closely when it comes to the charged particle multiplicity and
the subjet multiplicity. This indicates that the hybrid gives the best
total description of the hadronic $\mrm{e^{+}e^{-}}$ four-jet events at the  
$\mrm{Z^{0}}$ energy resonance.

A possible continuation of the work described in this paper is to expand
the hybrid to simulations of the full second order event 
q$\qbar$+q$\qbar$g+q$\qbar$gg+q$\qbar$q$\qbar$.
In such events the scale ambiguity in $\as$ affects the rate of 4-parton
events while the shape of these 4-parton events are unaffected. 
The first order matrix elements correction to parton showers has already 
been investigated by Michael H. Seymour \cite{ref:Seymour}.
Another much more difficult task is to apply the forced parton showers method 
to the description of hadron collisions.

\subsection*{Acknowledgment}
I would like to thank my supervisor T. Sj\"{o}strand for excellent
supervision.

\end{document}